\documentclass[a4paper,11pt]{article}
\usepackage{pos}

\newcommand{\be}{\begin{equation}}
	\newcommand{\ee}{\end{equation}}
\newcommand{\bea}{\begin{array}}
	\newcommand{\ea}{\end{array}}
\newcommand{\beqa}{\begin{eqnarray}}
	\newcommand{\eeqa}{\end{eqnarray}}
\newcommand{\nn}{\nonumber}

\usepackage{physics}
\newcommand{\del}{\partial}
\usepackage{bm}
\usepackage{bbm}
\usepackage{multirow}
\usepackage{mathtools}
\usepackage{dsfont}
\usepackage{float}
\usepackage{caption}
\usepackage{subcaption}

\title{Chaos in Matrix Gauge Theories with Massive Deformations}

\author[a]{K. Ba\c{s}kan}
\author*[a]{S. K\"{u}rk\c{c}\"{u}o\v{g}lu}
\author[a]{O. Oktay}
\author[a,b]{C. Ta\c{s}cı}

\affiliation[a]{Middle East Technical University, Department of Physics,\\Dumlupinar Boulevard, 06800, Ankara, Turkey}

\affiliation[b]{The Graduate Center, City University of New York, \\
 365 Fifth Ave, New York, NY 10016, U.S.A.}

\emailAdd{kagan.baskan@metu.edu.tr}
\emailAdd{kseckin@metu.edu.tr}
\emailAdd{oktay24005@gmail.com}
\emailAdd{ctasci@gradcenter.cuny.edu}

\abstract{Starting from an $SU(N)$ matrix quantum mechanics model with massive deformation terms and by introducing an ansatz configuration involving fuzzy four- and two-spheres with collective time dependence, we obtain a family of effective Hamiltonians, $H_n \,, (N = \frac{1}{6}(n+1)(n+2)(n+3))$ and examine their emerging chaotic dynamics.  Through numerical work, we model the variation of the largest Lyapunov exponents as a function of the energy and find that they vary either as $\propto (E-(E_n)_F)^{1/4} $ or $\propto E^{1/4}$, where $(E_n)_F$ stand for the energies of the unstable fixed points of the phase space. We use our results to put upper bounds on the temperature above which the Lyapunov exponents comply with the Maldacena-Shenker-Stanford (MSS) bound, $2 \pi T $, and below which it will eventually be violated.}

\FullConference{%
  Corfu Summer Institute 2021 "School and Workshops on Elementary Particle Physics and Gravity"\\
  29 August - 9 October 2021\\
  Corfu, Greece
}

\tableofcontents

\begin{document}
\maketitle

\section{Introduction}

Interest in exploring various aspects of chaotic dynamics emerging from Yang-Mills (YM) matrix models has been going on well over a decade now \cite{Sekino:2008he,  Asplund:2011qj,  Shenker:2013pqa, Gur-Ari:2015rcq, Berenstein:2016zgj, Maldacena:2015waa, Maldacena:2016hyu, Aoki:2015uha, Asano:2015eha, Berkowitz:2016jlq, Buividovich:2017kfk, Buividovich:2018scl, Coskun:2018wmz, Baskan:2019qsb}. It is well-known that investigations on the chaotic dynamics of Yang-Mills theories goes back to early 80s \cite{Matinyan:1981dj, Savvidy:1982wx, Savvidy:1982jk}. In the context of the Banks-Fischler-Shenker-Susskind (BFSS) model \cite{Banks:1996vh} the earliest work is due to Arefeva et. al. \cite{Arefeva:1997oyf}. A large portion of the recent work on the subject is motivated by a result due Maldacena-Shenker-Stanford (MSS) \cite{Maldacena:2015waa}, which states that the largest Lyapunov exponent (which is a measure of chaos in both classical and quantum mechanical systems) for quantum chaotic dynamics is controlled by a temperature dependent bound and given by $\lambda_L \leq 2\pi T$. It is conjectured that systems which are holographically dual to the black holes saturate this bound and therefore they can be deemed to be maximally chaotic. One such example is the Sachdev-Ye-Kitaev (SYK) matrix model \cite{Maldacena:2016hyu}, and it is expected to be so for other matrix models which have a holographic dual such as the BFSS \cite{Banks:1996vh} matrix quantum mechanics. The latter and its massive deformation Berenstein-Maldacena-Nastase (BMN) model \cite{Berenstein:2002jq} are supersymmetric $SU(N)$ gauge theories, describing the dynamics of $N$-coincident $D0$-branes in the flat and spherical backgrounds, respectively \cite{Banks:1996vh, deWit:1988wri, Itzhaki:1998dd,  Berenstein:2002jq, Dasgupta:2002hx, Ydri:2017ncg,Ydri:2016dmy}. Gravity dual of the BFSS model describes a phase in which $D0$-branes form a so called black-brane, i.e. a string theoretical black hole \cite{Ydri:2017ncg,Ydri:2016dmy,Kiritsis:2007zza}. 

As noted in \cite{Gur-Ari:2015rcq}, classical dynamics of the BFSS model serves as a good approximation of the quantum theory in the high temperature limit. Although this regime is distinguished from that in which the gravity dual is obtained, numerical studies conducted in \cite{Anagnostopoulos:2007fw, Catterall:2008yz} give no indication of an occurrence of a phase transition between the low and high temperature limits, which makes it plausible to expect that features like fast scrambling \cite{Sekino:2008he} of black holes in the gravity dual could survive at the high temperature limit too. Also, numerical results supporting fast thermalization is obtained for the BMN matrix model in \cite{Asplund:2011qj}. In \cite{Gur-Ari:2015rcq} classical chaotic dynamics of the BFSS models is studied and it is found that the largest Lyapunov exponent is given as $\lambda_L = 0.2924(3) T^{1/4}$ (setting $\lambda_{'t \, Hooft}$ to unity after properly scaling all the dimensionful quantities). Since the classical theory is only good in describing the dynamics in the high temperature regime, it is to be expected that it breaks down at sufficiently low temperatures. Indeed, it is readily seen that the MSS bound is violated for $T < T_c \approx 0.015$ while it remains parametrically smaller than $2 \pi T$ for $T > T_c$.  Let us also note that authors of \cite{Asano:2015eha} considered fuzzy spheres at matrix levels $N=2,3$ to exhibit the presence of chaotic dynamics., while in \cite{Coskun:2018wmz} some of us focused on a Yang-Mills five matrix models with a mass term and considering equivariant fluctuations around its fuzzy four sphere \cite{Castelino:1997rv,Kimura:2002nq,Ydri:2016dmy} solutions examined the chaotic dynamics of a family of effective models by computing their Lyapunov exponents. 

In \cite{Baskan:2019qsb}, we have studied chaos in massive deformations of the $SU(N)$ Yang-Mills gauge theories in $0+1$-dimensions. In this talk we report on this work. The model we are interested in has the same matrix content as that of the bosonic part of the BFSS model but it involves two separate massive deformation terms, which in two distinct limits lead to the solutions of the classical equations of motion in the form of fuzzy two- or four-spheres. Since non-trivial $D0$-brane dynamics leading to chaos require non-commutativity \cite{Aoki:2015uha}, fuzzy spheres described by non-commuting matrices may serve as good candidates for background geometries to probe chaotic behaviour in YM matrix models. Introducing an ansatz configuration involving fuzzy four- and two-spheres with collective time dependence, we obtain a family of effective Hamiltonians, $H_n \,, (N = \frac{1}{6}(n+1)(n+2)(n+3))$ and examine their emerging chaotic dynamics. Subsequently, we model the variation of the largest Lyapunov exponents as a function of the energy, and find that they vary either as $\propto (E -(E_n)_F)^{1/4} $ or $\propto E^{1/4}$, where $(E_n)_F$ stand for the energies of the unstable fixed points of the phase space. Using the virial and equipartition theorems we derive inequalities that relate energy and temperature and put upper bounds on the temperature above which the Lyapunov exponents comply with the Maldacena-Shenker-Stanford (MSS) bound $2 \pi T $, and below which it will eventually be violated.

\section{Yang-Mills Matrix Models with Double Mass Deformation}

In this section, we introduce a Yang-Mills (YM) matrix models with two distinct mass deformation terms, which may be contemplated as a double mass deformation of the bosonic part of the BFSS model. The action for this model can be written as 
\be
S_{YMM} = N \int dt \, L_{YMM} := N \int dt \, Tr \left ( \frac{1}{2}(D_0 X_I)^2 + \frac{1}{4} [X_I, X_J]^2 - \frac{1}{2} \mu_1^2 X_a^2 - \frac{1}{2} \mu_2^2 X_i^2 \right) \,, 
\label{YMDM}
\ee
where the indices $I,J=1,..,9$, while $a$ and $i$ take on the values $a = 1,..,5$ and $i = 6,7,8$, respectively. $X_I$ are $N \times N$ Hermitian matrices transforming under the adjoint representation of $U(N)$ as
\be
X_I \rightarrow U^{\dagger} X_I U \,, \quad U \in U(N) \,,
\ee
and the covariant derivatives are given by
\be
D_0 {X}_I=\del_0 X_I - i \lbrack A_0, X_I \rbrack \,.
\ee
$A_0$ is the $U(N)$ gauge field transforming as
\be
A_0 \rightarrow U^{\dagger} A_0 U + i U^{\dagger} \del_0 U. 
\ee
$S_{YMM}$ is invariant under the $U(N)$ gauge symmetry. In (\ref{YMDM}) the terms proportional to $\mu_1^2$ and $\mu_2^2$ are the quadratic deformations terms.
In the absence of the latter the action reduces to the bosonic sector of the BFSS model and is invariant under rigid $SO(9)$ rotations
\be
X_I \rightarrow X_I^\prime = R_{IJ} X_J \,, \quad R \in SO(9) \,,
\ee
of the matrices $X_I$ among themselves. Massive deformation terms breaks this global symmetry to $SO(5) \times SO(3) \times {\mathbb Z_2}^{\otimes 9}$, where the discrete group factor is due to the $X_I \rightarrow - X_I$ symmetry.

We will work in the 't Hooft limit, introducing $\lambda_{' t Hooft} = g^2 N$ with $N$ taken large and $g^2$ small, so that the 't Hooft coupling $\lambda_{' t Hooft}$ is held fixed. $\lambda_{' t Hooft}$ has units of $\mbox{Length}^{-3}$ and without loss of generality it can be set to unity by scaling all the dimensionful quantities in the action in units of $\lambda^{1/3}$.  The action in (\ref{YMDM}) is already written in the 't Hooft limit with $\lambda_{' t Hooft}$ set to unity. It is always possible to restore back $\lambda_{' t Hooft}$ by the scalings $X_i \rightarrow \lambda^{-1/3} X_i$, $A_t\rightarrow \lambda^{-1/3} A_t$, $t \rightarrow \lambda^{1/3} t$, $\mu_i^2 \rightarrow \lambda^{-2/3} \mu_i^2$, if needed.

Let us consider the scaling transformation
\be 
X_I  \rightarrow  \rho^{-1} \, X_I  \,, \quad t \rightarrow \rho \, t \,, 
\label{st}
\ee
where $\rho$ is an arbitrary positive constant. Under this transformation the conjugate momenta $P_I : = N \partial_0 X_I$ scales as $P_I \rightarrow \rho^{-2} P_I$. In the massless limit the potential scales as $V |_{(\mu_1, \mu_2 = 0)} \rightarrow \rho^{-4} \, V |_{(\mu_1,\mu_2 = 0)}$, indicating that the energy scales as $E \rightarrow \rho^{-4} E $. Since the Lyapunov exponent has the dimensions of inverse time, we see that, it should scale as $\lambda_L \propto E^{1/4}$ in the massless limit. These considerations will be of use in the following sections.

Fixing the gauge to $A_0=0$ yields the equations of motion for the matrices $X_I$'s in the form 
\begin{subequations}
	\begin{align}
		\ddot{X_a} + \lbrack X_I , \lbrack X_I , X_a \rbrack \rbrack  + \mu_1^2 X_a &= 0 \,, \label{YMDMeomA}\\ 
		\ddot{X_i} + \lbrack X_I , \lbrack X_I , X_i \rbrack \rbrack +  \mu_2^2 X_i &= 0 \,, \label{YMDMeomB} \\
		\ddot{X_9} + \lbrack X_I , \lbrack X_I , X_9 \rbrack \rbrack &= 0 \label{YMDMeomC}\,,
	\end{align}
	\label{YMDMeom} 
\end{subequations} and they are supplemented by the equation of motion for the gauge field $A_0$, which is nothing but the Gauss law constraint 
\be
\lbrack X_I, P_I \rbrack = 0 \,.
\label{GaussLaw1}
\ee
This is a constraint on the matrices $X_I$ and in the quantum theory it enforces that the physical states are $SU(N)$-singlets.

BMN matrix model \cite{Berenstein:2002jq}, is a particular massive deformation of the BFSS model, which preserves the maximal amount of the supersymmetry and it has the fuzzy two-spheres and their direct sums as possible vacuum configurations.  In what follows our focus will be directed at the emergent chaotic dynamics from the Yang-Mills matrix models. For this purpose, it will prove useful have YM  matrix models that could carry not only fuzzy two-sphere configurations but also higher dimensional fuzzy spheres, in particular a fuzzy four-sphere, as possible vacuum solutions. The specific massive deformation introduced (\ref{YMDM}) is distinct from that of the BMN model and serves precisely for this purpose, since in two distinct limiting cases, the equations of motion 	\eqref{YMDMeom}  can be solved either with fuzzy two-sphere or fuzzy four-sphere configurations. These are as follows:

For, $X_i = 0 = X_9$, (\ref{YMDMeomB}) and (\ref{YMDMeomC}) are satisfied identically, while (\ref{YMDMeomA}) takes the form 
\be
\ddot{X_a} + \lbrack X_b , \lbrack X_b , X_a \rbrack \rbrack  + \mu_1^2 X_a = 0 \,,
\label{foursphereeom} 
\ee
which is satisfied by the fuzzy four-sphere configurations $X_a \equiv Y_a$ for $\mu_1^2=-16$. The latter is described by the $N \times N$ matrices $Y_a$ carrying the $(0,n)$ UIR of $SO(5)$ with $N = \frac{1}{6}(n+1)(n+2)(n+3)$, i.e. the dimension of the UIR $(0,n)$. They satisfy the defining properties
\beqa
Y_a Y_a  &=&  \frac{1}{4}n(n+4) {\mathds{1}}_N \,, \nn \\
\epsilon^{a b c d e} Y_a Y_b Y_c Y_d &=& (n+2) Y_e \,, 
\label{fuzzyS4}     
\eeqa
A quick summary for the construction of the fuzzy four-sphere is given in the appendix of \cite{Coskun:2018wmz}, while more detailed accounts may be found in the original papers \cite{Castelino:1997rv,Kimura:2002nq,Ydri:2016dmy}.

For, $X_a = 0 = X_9$ on the contrary, the only remaining non-trivial equation of motion is
\be
\ddot{X_i} + \lbrack X_j , \lbrack X_j , X_i \rbrack \rbrack  + \mu_2^2 X_i = 0 \,,
\label{twosphereeom}
\ee
and for $\mu_2^2 = - 2$ it is solved by fuzzy two-sphere configurations $X_i \equiv Z_i$ or their direct sums. In this case, $Z_i$ are $N \times N$ matrices carrying the spin $j = \frac{N-1}{2}$ UIR of $SO(3) \approx SU(2)$. A detailed account on the fuzzy two-spheres and their applications is \cite{Balachandran:2005} which also provides an extensive list of references to the original literature.

In the ensuing section we consider an ansatz configuration involving fuzzy two- and four- spheres with collective time dependence. The latter implies that the Gauss law constraint (\ref{GaussLaw1}) is easily satisfied as we shall shortly see. Tracing over the fuzzy two- and four-sphere configurations yields reduced models with only four phase space degrees of freedom and their dynamics is readily accessible for numerical study.  

\subsection{Ansatz and the Effective Action}

A simple, yet non-trivial configuration involving fuzzy four- and two-sphere matrices with collective time-dependence is 
\be
X_a =  r(t) \, Y_a \,, \quad X_i =  y(t) \, Z_i \,, \quad X_9 = 0 \,,
\label{Anstz1}    
\ee 
where $r(t)$ and $y(t)$ are real functions of time. In this ansatz, we consider a single spin-$j = \frac{N-1}{2}$ IRR of $SU(2)$ as the fuzzy $S^2$ configuration, Although $Z_i$ exists at every matrix level, $Y_a$ do not. Fuzzy four spheres exists at the matrix levels $4,10,20 \cdots$ as given by the dimension $N = \frac{1}{6}(n+1)(n+2)(n+3)$ of the IRR $(0,n)$ of $SO(5)$. Accordingly we take the fuzzy two spheres at the matching matrix levels with that of the fuzzy four sphere. In what follows, we will focus on two distinct set of mass values which are $(\mu_1^2=-16, \mu_2^2=-2)$ and $(\mu_1^2=-8, \mu_2^2=1)$. The former are the mass values required for the static solutions of fuzzy two- or four-spheres when either the $X_a$'s or the $Z_i$'s are set to zero, respectively, while the latter is an possible example, among many, of mass values leading to a single trivial fixed point for the reduced Hamiltonians. which will shortly follow. 

Substituting the configuration (\ref{Anstz1})  in the action (\ref{YMDM}) and tracing oover the fuzzy four- and two-sphere matrices at the matrix levels $N = \frac{1}{6}(n+1)(n+2)(n+3)$ for $n=1,2,\cdots\,, 6$, we obtain the family of Lagrangians of the reduced models in the form  
\be
L_n = N^2( c_{1} \dot{r}^2 +  c_{2}  \dot{y}^2  - 8  c_{1}  r^4 - c_{2}  y^4 -  c_{1}  \mu_1^2 r^2 -  c_{2}  \mu_2^2 y^2 -  c_{3} r^2y^2) \,,
\label{Lalp}
\ee
where the coefficients $c_{\beta} = c_{\beta}(n)$ ($\beta=1,2,3$) depend on $n$ and their values (given up to one digit after the decimal point at most) for $n=1,2,\cdots\,, 6$ are listed in the table \ref{table:table1}. We suppress the label $n$ of the coefficients $c_{\beta}(n)$ in (\ref{Lalp}) in order not to clutter the notation. 
\begin{table}[!htb]
	\centering
	\begin{tabular}{ | c | c | c | c | c | c | c |}
		\cline{2-7}
		\multicolumn{1}{c |}{} & $n=1$ & $n=2$ & $n=3$ & $n=4$ & $n=5$ & $n=6$ \\ \hline 
		$c_{1}$ &$2.5$  & $6$ & $10.5$  & $16$ &$22.5$  &$30$     \\ \hline 
		$c_{2}$ &$1.9$  & $12.4$  &$49.9$  &$153$ &$391.9$ &$881.9$   \\ \hline 
		$c_{3}$ &$21$ & $207.7$ &$1080$ &$3970$ &$11691$ &$29493$ \\ \hline
	\end{tabular}
	\caption{Numerical values of coefficients $c_\beta(n)$.}
	\label{table:table1}
\end{table}

The corresponding Hamiltonian is 
\beqa
H_{n}(r,y,p_r,p_y) &=& \dfrac{{p_r}^2}{4c_{1}N^2} + \dfrac{{p_y}^2}{4c_{2}N^2} + N^2(8  c_{1} \, r^4 + c_{2} \, y^4 +  c_{1} \, \mu_1^2 r^2 +  c_{2} \, \mu_2^2 y^2 + c_{3} \, r^2y^2) \,, \nn \\
&= :& \dfrac{{p_r}^2}{4c_{1}N^2} + \dfrac{{p_y}^2}{4c_{2}N^2}  + N^2 V_n(r,y) \,,
\label{Hn1}
\eeqa
where $V_n(r,y)$ stands for the potential and defined by the relevant terms in the first line of \eqref{Hn1}. Hamilton's equations of motion take the form 
\begin{subequations}
	\begin{align}
		\dot{r} - \dfrac{p_r}{2 c_{1}N^2} = 0	\,, \quad \quad & \dot{y} - \dfrac{p_y}{2 c_{2}N^2} = 0 \,, \label{Heom1} \\
		\dot{p_r} + N^2( 32  c_{1} r^3 + 2  c_{1} \mu_1^2 r + 2  c_{3} r  y^2)  = 0 \,, \quad \quad  & \dot{p_y} + N^2(4 c_{2} y^3 + 2 c_{2} \mu_2^2 y + 2 c_{3} r^2 y)  = 0 	\,. \label{Heom2}
	\end{align}
	\label{Heom} 
\end{subequations}
Case of $n=1$ does not lead to Chaotic dynamics and will not be considered in the ensuing sections. Details pertaining to this case may be found in \cite{Baskan:2019qsb}.

 Let us also note in passing that in the massless limit  $(\mu_1 \,, \mu_2) \rightarrow (0,0)$, we have $H_n \rightarrow \rho^{-4} H_n$ under the scaling $(r \,, y)  \rightarrow (\rho^{-1} \, r  \,, \rho^{-1} \, y)$ and $t \rightarrow \rho \, t $, as can be readily expected in view of the discussion given in the previous section.

\subsection{Fixed Points and Stability Analysis}

To investigate the dynamics of the models described by the Hamiltonian's $H_n$, it is useful to start by determining the fixed points in the phase space and subsequently inspect their stability at the linear order. Fixed points in the phase space can be determined by solving the equations \cite{Percival,Ott,Hilborn,Campbell}, 
\be
(\dot{r}, \dot{y}, \dot{p_r}, \dot{p_y}) \equiv (0,0,0,0) \,.
\label{fpcond} 
\ee
Using (\ref{fpcond}) in (\ref{Heom}) leads to four algebraic equations, two of which are immediately solved by $(p_r, p_y) \equiv (0,0)$. This means that all the fixed points are on the $(p_r, p_y) \equiv (0,0)$ plane in the phase space. Taking the mass parameter values as $\mu_1^{2}=-16$ \& $\mu_2^{2}=-2$, the remaining two equations are
\beqa
- 32  c_{1} r^3 +32  c_{1}  r - 2  c_{3} r y^2 & = & 0 \,, \nn \\
- 4 c_{2} y^3 +4 c_{2} y - 2 c_{3} r^2 y & = & 0 \,,
\label{fpeqns1}
\eeqa
Only real solutions of (\ref{fpeqns1}) are physically acceptable. The fixed points are
\begin{multline}
	(r,y,p_r,p_y) \equiv \lbrace (0,0,0,0) , (  \pm1,0,0,0) \,, ( 0, \pm 1,0,0) ) \,,\\
	(\pm h_1(n), \pm h_2(n),0,0) \,, (\pm h_1(n), \mp h_2(n),0,0)   \rbrace \,, 
	\label{real2}
\end{multline}
where $h_1$ and $h_2$ are given in terms of $c_\beta$ as
\be
h_1=- \sqrt{2} i \frac{\sqrt{-c_2c_3+ 16 c_1c_2}}{\sqrt{c_3^2-32c_1c_2}} \,, \quad h_2= - 4 i \frac{\sqrt{2c_1c_2-c_1c_3}}{\sqrt{c_3^2-32c_1c_2}} \,.
\ee
with the numerical values are presented in the table \ref{table:table2} below.
\begin{table}[!htb]
	\centering
	\begin{tabular}{ | c | c | c | c | c | c |}
		\cline{2-6}
		\multicolumn{1}{c |}{} & $n=2$ & $n=3$ & $n=4$ & $n=5$&$n=6$ \\ \hline 
		$h_1$  & $0.26$ & $0.28$  & $0.27$ &$0.26$ &$0.24$  \\ \hline 
		$h_2$ & $0.6$  &$0.38$  &$0.25$ &$0.17$ &$0.12$  \\ \hline 
	\end{tabular}
	\caption{Numerical values of $h_1$ and $h_2$.}
	\label{table:table2}
\end{table}
Energies at these fixed points are readily evaluated and they are  
\be
E_F(0, 0, 0, 0)= 0 \,,\quad E_F(\pm 1,0,0,0) = -8 N^2 c_1 \,, \quad E_F(0,\pm 1,0,0) = - N^2 c_2 \,, 
\label{fpenergies1}
\ee
while the values of $\frac{E_F(\pm h_1(n), \pm (\mp) h_2(n), 0, 0)}{N^2}$ are presented in \ref{table:table3}. Same table also lists minimum values of the potentials $V_n$ for ease in comparison. 
\begin{table}[!htb]
	\centering
	\begin{tabular}{|c| c | c | c | c | c | }
		\cline{2-6}
		\multicolumn{1}{c|}{} & $n=2$ & $n=3$ & $n=4$ & $n=5$&$n=6$ \\ \hline 
		$\frac{E_F(\pm h_1(n), \pm (\mp) h_2(n), 0, 0)}{N^2}$ & $-8.6$ & $-13.8$ & $-18.4$  &$-23$ &$-27.7$ \\ \hline
		$\frac{V_{(n), min}}{N^2}$  & $-48$ & $-84$ & $-153$  &$-391.9$ &$-881.9$ \\ \hline
	\end{tabular}
	\caption{Numerical values for $E_F/N^2$ at the critical points $(\pm h_1(n), \pm(\mp)h_2(n), 0, 0)$ and the minimum values of $V_n$.}
	\label{table:table3}
\end{table}
Let us note that the critical points of the potential $V_n$ are also determined by the solutions of (\ref{fpeqns1}) . From the eigenvalues of the matrix $\frac{\partial^2 V_n}{\partial g_i \partial g_j}$,  (with the notation $(g_1, g_2) \equiv (r,y)$), we see that $(\pm 1,0)$ and $(0 \,,\pm 1)$ are local minima, $(0,0)$ is a local maximum, and  $(\pm h_1(n), \pm h_2(n))$, $(\pm h_1(n), \mp h_2(n))$ are all saddle points. Evaluating $V_n$ at the local minima, we find $V_n(\pm 1,0) = -8 c_1$ and $V_n(0, \pm 1) = - c_2$. These give the absolute minimum of $V_n$ at $(\pm 1,0)$ for $n=1,2,3$ and at $(0, \pm 1)$ for $n =4,5,6$.  We may also note that minimum of $V_n$ are negative in general. This is expected, due to the presence of the massive deformation terms. 

A first order stability analysis around the fixed points of $H_n$ can be readily given. Together with the Lyapunov spectrum that will be determined in the next section, this analysis will allow us to comment on the onset of chaos, i.e.  to estimate the energies at and above which chaotic orbits will start to dominate the assocaited phase spaces of $H_n$. Using the notation
\be
(g_1, g_2, g_3, g_4) \equiv (r, y, p_r, p_y) \,,
\label{psc}
\ee
the Jacobian matrix may be given as
\be
J_{\alpha \beta} \equiv \frac{\partial \dot{g}_\alpha}{\partial g_\beta} \,,
\ee
Eigenvalues of $J_{\alpha \beta}$ at the fixed points are easily evaluated and allows us to decide on their stability. General criterion for the latter states that \cite{Percival,Ott,Hilborn,Campbell} a fixed point is stable if all the real eigenvalues of the Jacobian are negative, unstable if the Jacobian has at least one real positive eigenvalue, and of borderline type if all the eigenvalues are purely imaginary\footnote{That is, the first order analysis is inconclusive and higher order considerations are necessary to decide if the system is stable or unstable at such a fixed point.}. Accordingly, we find that $(0,0,0,0)$ and $(h_1(n),h_2(n),0,0)$ are all unstable fixed points, while we find that $(\pm1,0,0,0)$ and $(0,\pm 1, 0, 0)$ are borderline type. We are not going to explore the structure of the latter, since we expect that their impact on the chaotic dynamics be rather small compared to those of the unstable fixed points. Our results on the Lyapunov spectrum presented in the next section corroborates with this expectation. 

\section{Chaotic Dynamics}

\subsection{Dependence of the Largest Lyapunov Exponent on Energy}

In order to probe the presence and analyze the structure of chaotic dynamics of the models described by the Hamiltonians $H_n$, we examine their Lyapunov exponents. The latter, and in particular the largest, Lyapunov exponent give a measure on the exponential growth in perturbations and therefore serve as a quantitative means to establish the presence of chaos in a dynamical system \cite{Ott,Hilborn,Campbell}. In order to obtain the Lyapunov exponents for our models we run a Matlab code, which numerically solves the Hamilton's equations of motion given in (\ref{Heom}) for all $H_n$ ($2 \leq n \leq 6$) at several different values of the energy. We run the code $40$ times with randomly selected initial conditions satisfying a given energy and calculate the mean of the time series from all runs for each of the Lyapunov exponents at each value of $n$. Details in this regard and a simple approach we developed to give certain effectiveness to the random initial condition selection process are presented in \cite{Baskan:2019qsb}. Here we directly focus on the profile of the Largest Lyapunov exponents as a function of the $E/N^2$.

\begin{figure}[!htb]
	\centering
	\begin{subfigure}[!htb]{.32\textwidth}
		\centering
		\includegraphics[width=1\linewidth]{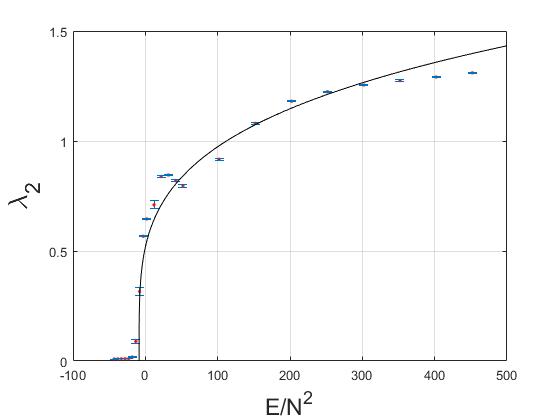}  
		\caption{$\lambda_2$ vs. $E/N^2$}
		\label{fig:fig2b}
	\end{subfigure}	
	\begin{subfigure}[!htb]{.32\textwidth}
		\centering
		\includegraphics[width=1\linewidth]{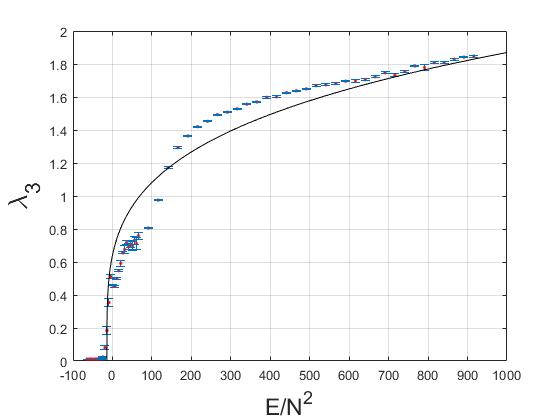}  
		\caption{$\lambda_3$ vs. $E/N^2$}
		\label{fig:fig2c}
	\end{subfigure}	
	\begin{subfigure}[!htb]{.32\textwidth}
		\centering
		\includegraphics[width=1\linewidth]{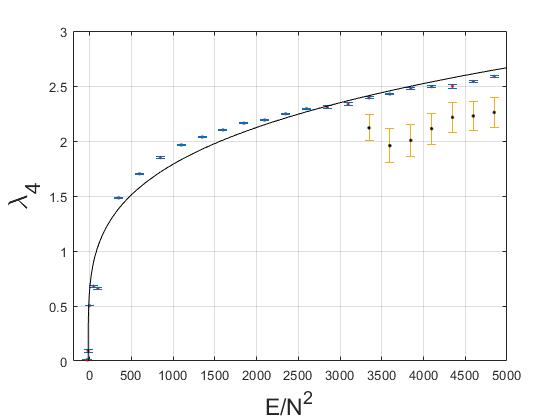}  
		\caption{$\lambda_4$ vs. $E/N^2$}
		\label{fig:fig2d}
	\end{subfigure}		
	\begin{subfigure}[!htb]{.32\textwidth}
		\centering
		\includegraphics[width= 1\linewidth]{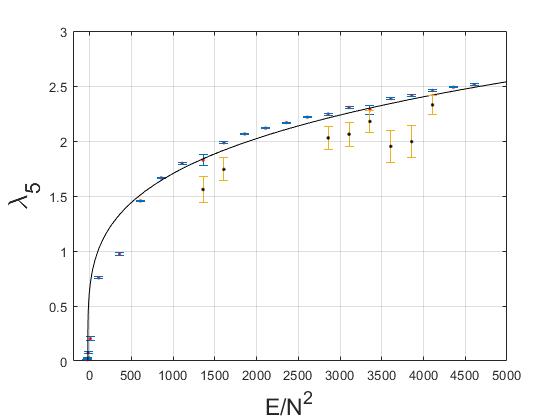}  
		\caption{$\lambda_5$ vs. $E/N^2$}
		\label{fig:fig2e}
	\end{subfigure}	
	\begin{subfigure}[!htb]{.32\textwidth}
		\centering
		\includegraphics[width=1\linewidth]{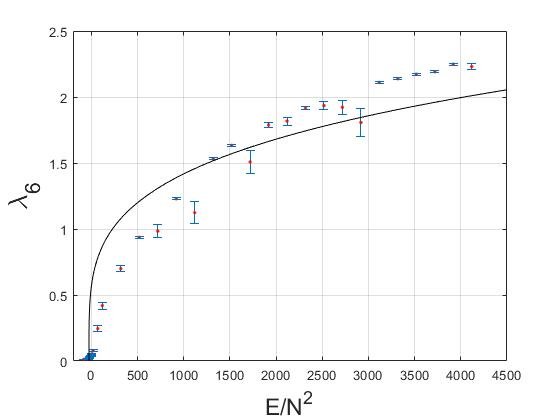}  
		\caption{$\lambda_6$ vs. $E/N^2$}
		\label{fig:fig2f}
	\end{subfigure}		
	\caption{MLLE vs. $E/N^2$ for $\mu_1^2=-16$ and $\mu_2^2 =-2$}
	\label{fig:fig2}
\end{figure}

In order to see the dependence of the mean largest Lyapunov exponent, $\lambda_n$, to energy, we compute the latter at a large span of energy values, which suits best to observe the onset and development of the chaotic dynamics. The energies determined for the unstable fixed points in the previous section are of central importance here. From set of data presented in figure \ref{fig:fig2}, we see that chaotic dynamics starts to develop only after the energy of systems exceeds the unstable fixed point energies $(E_n)_F$ of the models at the fixed points $(\pm h_1(n), \pm (\mp) h_2(n), 0, 0)$. As we keep on increasing the energy, $\lambda_n$ tend to grow as is easily observed from the figure \ref{fig:fig2}. Error bars at each data point is found by evaluating mean square error using the average and the individual largest Lyapunov exponent values for each of the $40$ runs of the code.

In view of the expectation that $\lambda_n \propto E^{1/4}$ due to the scaling symmetry of the models $H_n$ in the massless limit and the fact that the reduced models do not develop any appreciable chaos up until the critical energies $(E_n)_F < 0$ are well exceeded. We examine the best fit curves of the form 
\be
\lambda_n = \alpha_n \left (\frac{E}{N^2} - \frac{(E_n)_F}{N^2} \right)^{1/4} \,,
\label{lvE}
\ee
to the $\lambda_n$ versus $\frac{E}{N^2}$ data in figure \ref{fig:fig2}.  Let us also note that at level $n$, $(E_n)_F$ is determined by the values of the masses $\mu_1^2$, $\mu_2^2$, and therefore (\ref{lvE}) involves the dependence of $\lambda_n$ on these additional dimensionful parameters through $(E_n)_F$. This is in contrast to the pure BFSS model, whose only dimensionful parameter is $\lambda_{'t Hooft}$. Coefficients $\alpha_n$ for (\ref{lvE}) are listed in table \ref{table:fitvalue1} below:
\begin{table}[!htb]
	\centering
	\begin{tabular}{|c|c|c|c|}
		\hline
		$n$ &  $\alpha_{n}$ &$T_c$ & $\beta_{n}$ \\ \hline 
		$2$ & {0.3017} & 0.0832 & 0.4567 \\ \hline  
		$3$	& {0.3313} & 0.1029 & 0.5015 \\ \hline
		$4$	& {0.3168} & 0.1046 & 0.4795 \\ \hline
		$5$	& {0.3016} & 0.1045 & 0.4565 \\ \hline
		$6$	& {0.2505} & 0.0911 & 0.3791 \\ \hline
	\end{tabular}
	\caption{$\alpha_n$ values for the best fit curves (\ref{lvE}), Upper bounds for $T_c$ and $\beta_n$ values for the inequality (\ref{lvT}).}
	\label{table:fitvalue1}
\end{table}

A number of comments are in order. We see that the $\alpha_n$ values are close to each other for $2\leq n \leq 5$ and the quality of the fitting curves are quite good, while it is smaller for $n = 6$ and there is a visible decrease in the quality of the fitting curves. It may be argued that the improved quality of the fits for $n=4$ and $n=5$ are due to the more detailed analysis of the data. In fact, at these matrix levels, some of the randomly picked initial values lead to vanishing Lyapunov exponents for a range of the energy values and they correspond to quasi-periodic orbits in the phase space. For these range of the energies, we evaluated the mean $\lambda_n$ by excluding the initial conditions that lead to vanishing Lyapunov exponents. The results obtained in this manner are given in the plots in the figures \ref{fig:fig2d} and \ref{fig:fig2f} with blue error bars as with the rest of the data, while those points with error bars in yellow color represent the mean $\lambda_n$ including those initial conditions leading to vanishing Lyapunov exponents. The latter are kept in the same plots for comparison. Nevertheless, such an analysis is not suitable for $n=6$ due to the scattered distribution of the relatively lower Lyapunov exponent values to almost the entire energy range of interest. A more sophisticated initial condition selection procedure, for instance by confining to small hyper-volumes of phase space, and averaging the LLE data over such hyper-volumes may help to decrease fluctuations on the data and subsequently enhance the quality of the fits too, but this is out of the scope of our present work.\footnote{Approach of some of the LLE values to zero implies that the systems' development in time, starting from these initial conditions are of either periodic or quasi-periodic type and not chaotic. We note that the mean largest Lyapunov exponent is still quite large and also demonstrated in \cite{Baskan:2019qsb} that the sample Poincar\'{e} section taken at one of these energies are densely chaotic, showing no sign of Kolmogorov-Arnold-Moser(KAM) tori, which would have signaled the presence of appreciable amount of quasi-periodic orbits. Therefore, we think that such periodic or quasi-periodic orbits occur in considerably small regions of the phase space.} 

\subsection{Temperature Dependence of the Lyapunov Exponent}

Chaotic dynamics of the BFSS models was examined at the classical level in \cite{Gur-Ari:2015rcq} in a real time formalism\footnote{This is in contrast to several earlier as well as some recent studies on the BFSS model and its deformations mainly targeting to explore their phase structure, which resort to the imaginary time formalism with period $\beta$, the inverse temperature \cite{Kawahara:2006hs, Kawahara:2007fn, DelgadilloBlando:2007vx, DelgadilloBlando:2008vi, Asano:2018nol, Berkowitz:2018qhn}.}. There, the authors found that the largest Lyapunov exponent varies as a function of the temperature in the form $\lambda = 0.2924(3)(\lambda_{'t Hooft} T)^{1/4}$.  A simple dimensional analysis yields $\lambda$ in units of $\it{Length}^{-1}$ as it should, since $\lambda_{'t Hooft}$ and $T$ have the units $\it{Length}^{-3}$ and $\it{Length}^{-1}$, respectively. Let us note that, scaling transformation in \eqref{st} is exact for the BFSS model and hence the Lyapunov exponents should scale with energy as $\lambda \propto E^{1/4}$ and subsequently gets related to the temperature via the use of the virial and equipartition theorems. To be more precise, for a purely quartic potential the latter yields, $\langle K \rangle =2 \langle V \rangle  = \frac{1}{2} n_{d.o.f} T$ (In units $k_B=1$) and $E = \langle K \rangle + \langle V \rangle = \frac{3}{4} n_{d.o.f} T$, where $n_{d.o.f}$ stands for the total number of degress of freedom. For the BFSS model, this is evaluated to be \cite{Gur-Ari:2015rcq} , $n_{d.o.f} = 8(N^2-1)-36$ after properly accounting for constraints and symmetries. The result $\lambda = 0.2924 T^{1/4}$ (setting $\lambda_{'t Hooft}$ to unity) of \cite{Gur-Ari:2015rcq} is parametrically smaller than the MSS bound, $ \lambda \leq 2 \pi T$, on quantum chaos for sufficiently large $T$, but violates it below the temperature $(\frac{0.292}{2 \pi})^\frac{4}{3} = 0.015$. This is expected, since the classical theory is good in approximating the quantum dynamics only in the high temperature regime.

From the perspective and results outlined in the previous paragraph, we proceed with the following strategy to model the variation of the Lyapunov exponents, $\lambda_n$, with temperature. Due to the massive deformation terms we introduced in the matrix model, the potential is not a polynomial of homogeneous degree in the matrices $X_I$ any longer, and therefore the virial theorem does not allow us to express the kinetic ,$\langle K \rangle$, and the potential, $\langle V \rangle$, energies as a multiple of each other. However, expressing the Lagrangian in (\ref{YMDM}) as $L_{YMM}=K - V$ and applying the virial theorem, we find
\beqa
2 \langle K \rangle &=& 2 \langle V \rangle -  2 \, Tr \frac{1}{4} [X_I, X_J]^2 \,, \nn \\
&=&	 4 \langle V \rangle -  2 \, Tr \, \frac{1}{2} \mu_1^2 X_a^2 - 2 \, Tr \, \frac{1}{2} \mu_2^2 X_i^2 \,.
\eeqa
Here, the first equality implies that $\langle V \rangle < \langle K \rangle $ is always satisfied, since $Tr [X_I, X_J][X_I, X_J]^\dagger = - Tr [X_I, X_J]^2 \geq 0$ for Hermitian $X_I$. Second line implies that $2\langle  V  \rangle \leq  \langle K \rangle$ for $\mu_1^2 <0 $, $\mu_2^2 < 0$. Thus, we have $E \leq  \frac{3}{4}  n_{d.o.f} T $, where now $ n_{d.o.f} = 7(N^2-1)-28$, since $X_9$ is already set to zero in our model and this decreases the number of d.o.f accordingly. We may take $n_{d.o.f} \approx 7N^2$ at large $N$. The ansatz in \eqref{Anstz1} roughly describes $D0$-branes on the fuzzy two- and four-spheres with open string stretching between them, and the effective action we obtain after tracing model their collective dynamics. This suggests that, we may consider the energy of the reduced models as being due to the $\approx 7 N^2$ degrees of freedom of the matrix model and hence consider the inequality 
\be
\frac{E}{N^2} \leq \frac{21}{4} T \,.
\label{EvT}
\ee
In view of this inequality, we conclude that at zero temperature, $E \leq 0$, while the mean largest Lyapunov exponents are non-vanishing until the energies drop below or around $(E_n)_F$ as clearly seen from the data presented in figure \ref{fig:fig2}. This means that already at zero temperature the MSS bound $ \lambda_L \leq 2 \pi T$ is violated by these effective models. Using (\ref{EvT}) and (\ref{lvE}) together readily yields $T_c$ as the temperature saturating the inequality $\alpha_n \left (\frac{21}{4} T - \frac{(E_n)_F}{N^2} \right)^{1/4} \geq 2 \pi T$.  For $n=2\,, \cdots\,,6$, numerical values of $T_c$ are given in table \ref{table:fitvalue1} and they are roughly an order of magnitude larger than the critical temperature $0.015$ determined for the BFSS model in \cite{Gur-Ari:2015rcq}. These are the upper bounds on the temperature below which the MSS bound will eventually be violated. Due to \eqref{EvT}, we can expect that the temperatures at which this actually happens 
should be less than $T_c'$'s and therefore closer to the value determined for the BFSS model. At sufficiently high energies, we may estimate
\be
\lambda_n \approx \alpha_n \left( 1 - \frac{1}{4} \frac{(E_n)_F}{E}\right) \left (\frac{E}{N^2} \right)^{1/4}  \,, \quad \mbox{for}\, \quad  E \gg |(E_n)_F| > 0 \,,
\ee 
and therefore 
\be
\lambda_n \leq \beta_n  \, T^{1/4} \,, \quad \beta_n := \left(\frac{21}{4} \right )^{1/4} \alpha_n  \,,
\label{lvT}
\ee
which is strictly valid in the high temperature regime for non-vanishing values of $\lambda_n$ and is parametrically smaller than the MSS bound $2 \pi T$. Numerical values of $\beta_n$ are provided in table \ref{table:fitvalue1} for easy access.

An important issue that needs to be addressed is how to compute quantum corrections to the matrix model and to the reduced effective actions presented in this paper or more generally in the broader context of the BFSS and related matrix models. We remark on some recent developments employing real time techniques in the conclusions.

\subsection{Another Set of Mass Values}

We may consider if and how the results of the previous section could get altered if we work with values of $\mu_1^2$ and $\mu_2^2$ for which no fixed point of the type $(h_1(n), h_2(n),0,0)$ with negative $E_F$ is present but only $(0,0,0,0)$ is the unstable fixed point with zero energy at each level $n$. Here we do not plan to present an exhaustive discussion but simply confine ourselves to examine another choice for the mass values to serve the aforementioned purpose. In particular, we take $\mu_1^2= -8$ and $\mu_2^2 = 1$.  

With the Lagrangians $L_n$ and the Hamiltonians $H_n$ given in the form (\ref{Lalp}) and (\ref{Hn1}) and the corresponding Hamilton's equations given as in (\ref{Heom}), we find that the local minimum of the potential is at $(r,y) = (\pm\frac{1}{\sqrt{2}},0)$ and the local maximum is at $(r,y) = (0,0)$ with the corresponding energies $V_{(n),min} = - 2c_1 N^2$ and $0$, respectively. 

Fixed points of the phase space are
\be
(r,y,p_r,p_y) =  \lbrace (0,0,0,0)\,, (\pm\frac{1}{\sqrt{2}},0,0,0) \rbrace 
\ee    
with the corresponding energies 
\be
E_F(0,0,0,0) =0  \,, \quad E_F(\pm \frac{1}{\sqrt{2}}, 0,0,0) = - 2 c_1 N^2 = - n(n+4) N^2 \,.   
\ee
Linear stability analysis demonstrates that $(0,0,0,0)$ is the only unstable fixed point, while $(\pm \frac{1}{\sqrt{2}},0,0,0)$ are of the borderline type and does not play a role in the ensuing discussion.

Following the same steps of the numerical analysis for the Lyapunov spectrum outlined previously in section 3.1, we find that, in this case too, the models exhibit chaotic dynamics for $n=2,3,4,5,6$.The transition to chaos starts to happen around the fixed point energies, i.e. just above zero energy, as can be clearly seen from these figures. 

\begin{figure}[!htb]
	\centering
\begin{subfigure}[!htb]{.32\textwidth}
		\centering
		\includegraphics[width=1\linewidth]{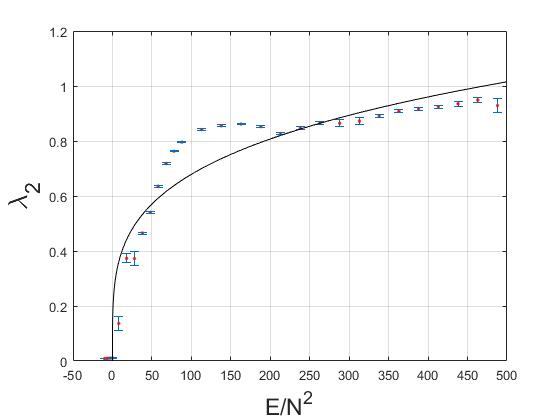}  
		\caption{$\lambda_2$ vs.$E/N^2$}
		\label{fig:fig4b}
	\end{subfigure}
	\begin{subfigure}[!htb]{.32\textwidth}
		\centering
		\includegraphics[width= 1\linewidth]{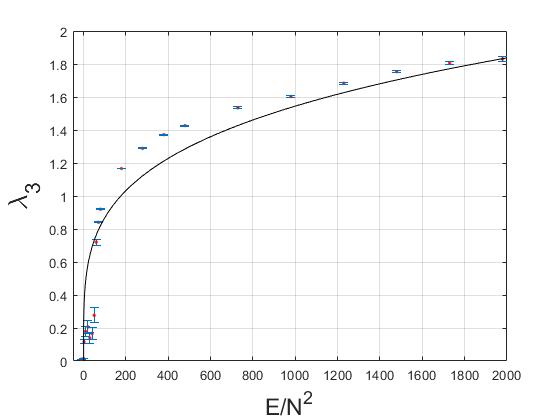}  
		\caption{$\lambda_3$ vs. $E/N^2$}
		\label{fig:fig4c}
	\end{subfigure}	
	\begin{subfigure}[!htb]{.32\textwidth}
		\centering
		\includegraphics[width=1\linewidth]{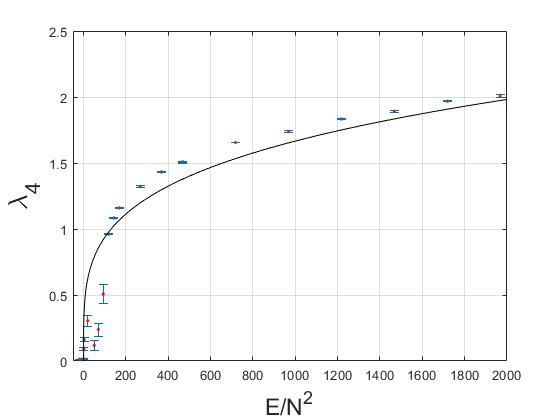}  
		\caption{$\lambda_4$ vs. $E/N^2$}
		\label{fig:fig4d}
	\end{subfigure}
	\begin{subfigure}[!htb]{.32\textwidth}
		\centering
		\includegraphics[width= 1\linewidth]{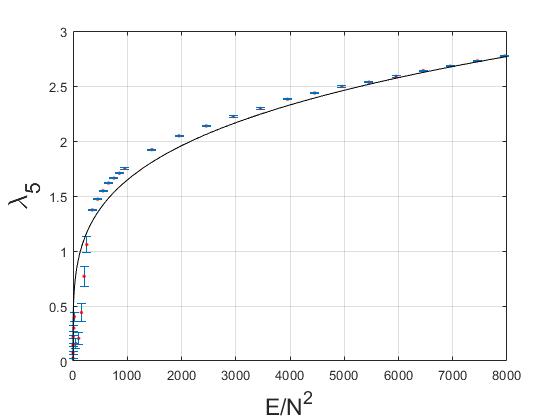}  
		\caption{$\lambda_5$ vs. $E/N^2$}
		\label{fig:fig4e}
	\end{subfigure}	
	\begin{subfigure}[!htb]{.32\textwidth}
		\centering
		\includegraphics[width=1\linewidth]{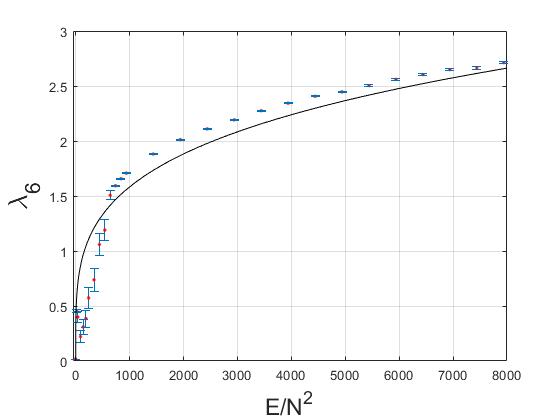}  
		\caption{$\lambda_6$ vs. $E/N^2$}
		\label{fig:fig4f}
	\end{subfigure}
	\caption{MLLE vs.$E/N^2$ for $\mu_1^2=-8$ and $\mu_2^2 = 1$}
	\label{fig:fig4}
\end{figure}
Since, the unstable fixed point energy is zero at each value of $n$, we consider fitting curves of the form 
\be
\lambda_n = {\tilde \alpha}_n \left ( \frac{E}{N^2} \right)^{1/4} \,.
\label{lvE2}
\ee
We see that fits given in figure \ref{fig:fig4} are quite good, while they predict somewhat lower values of $\lambda_n$ at larger energies, except at the level $n=2$. The latter case displays a sharper increase the Lyapunov exponent at intermediate energies, which is not captured well by the fit. ${\tilde \alpha}_n$ are provided in the table \ref{table:fitvalues2} and happen to be quite close to each other except for $\alpha_2$.  
\begin{table}[!htb]
	\centering
	\begin{tabular}{ | c | c | c |c|}
		\cline{1-4}
		$n$ &  ${\tilde \alpha}_n$ & $T_c$ & ${\tilde \beta}_n$ \\ \hline
		$2$ &$0.2147$ &$0.021$ & 0.349 \\ \hline
		$3$ &$0.2747$ &$0.029$ & 0.447 \\ \hline
		$4$ &$0.2963$ &$0.033$ & 0.481 \\ \hline
		$5$ &$0.2927$ &$0.032$ & 0.476 \\ \hline 
		$6$ &$0.2815$ &$0.030$ & 0.458 \\ \hline
	\end{tabular}
	\caption{${\tilde \alpha}_n$ values for the best fit curves (\ref{lvE2}), upper bounds for ${\tilde T}_c$ and ${\tilde \beta}_n$ values for the inequality (\ref{lvT2}).}
	\label{table:fitvalues2}
\end{table}
Since $\mu_2^2 = 1 >0$, virial theorem no longer implies $2 \langle V \rangle \leq \langle K \rangle $ in general, but $ \langle V \rangle \leq \langle K \rangle $ is still valid. Thus, we have $E \leq  n_{d.o.f} T $ and therefore, instead of (\ref{EvT}), we arrive at the inequality
\be
\frac{E}{N^2} \leq 7 T \,,
\label{EvT2}
\ee
Data in figure \ref{fig:fig4} show that all $\lambda_n \rightarrow 0$ as $E \rightarrow 0$. Thus, as opposed to the previous case with $(E_n)_F < 0$, there is no violation of the MSS bound at zero temperature.  Similar to the previous case, temperatures saturating the inequality ${\tilde \alpha}_n (7 T)^{1/4} \geq 2 \pi T$, give the ${\tilde T}_c$'s below which the MSS bound will eventually be violated.  Numerical values of these critical temperatures are given in table \ref{table:fitvalues2} and they are significantly lower than those listed in table \ref{table:fitvalue1} and closer to $0.015$ of the BFSS model. Finally, we see that 
\be
\lambda_n \leq {\tilde \beta}_n \, T^{1/4} \,, \quad{\tilde \beta}_n := 7 ^{1/4} {\tilde \alpha}_n  \,,
\label{lvT2}
\ee
with ${\tilde \beta}_n$ values provided in table \ref{table:fitvalues2}, and they are all parametrically lower than $2 \pi T$ at temperatures above ${\tilde T}_c$.

\section{Conclusions and Outlook}

In this talk, we have focused on to describe some of the essential features of our work in \cite{Baskan:2019qsb}. Interested readers may find the details as well as some other aspects in that article. To briefly summarize, here we have studied the emergence of chaotic dynamics in a YM theory with the same matrix content as that of the BFSS model. For this purpose, we have used an ansatz configuration in the form of fuzzy two- and four- spheres matrices with collective time dependence and obtained a family of models described by the effective Hamiltonians $H_n$. Subsequent numerical analysis allowed us to compute the Lyapunov exponents as well as to inspect how the largest Lyapunov exponent vary with energy. We found that this variation fits well with $\lambda_n \propto (E-(E_n)_F)^{1/4}$ or $\lambda_n \propto E^{1/4}$, in accord with the power law behavior expected on the grounds of the scaling symmetry of the model in the massless limit. Deriving inequalities that relate energy and temperature by exploiting the virial and equipartition theorems, we have shown that for $H_n$ at mass values $\mu_1^2=-16$ and $\mu_2^2 = -2$, the MSS bound will eventually be violated at temperatures below the $T_c$'s listed in the table, this is addition to the violation at zero temperature, the latter being due to the presence of an unstable fixed point at each $n$ in the phase space, whose energy $(E_n)_F$, although being well above the minimum of the potential is still negative, while $\lambda_n$ remains non-zero around $(E_n)_F$ and hence at zero temperature due to (\ref{EvT}). Similar results are also reached for the models with mass-squared values $\mu_1^2=-8$ and $\mu_2^2 = 1$. However, in this case, the only unstable fixed point of the phase space is $(0,0,0,0)$ with zero energy at each level $n$, implying $\lambda_n \rightarrow 0$ for $E \rightarrow 0$ and hence there is no violation of the MSS bound at zero temperature as the distribution of data in figure  \ref{fig:fig4} also confirms. Critical temperatures ${\tilde T}_c$ listed in the table \ref{table:fitvalue1} for this case are quite smaller than $T_c$ and closer to the value obtained in \cite{Gur-Ari:2015rcq} for the BFSS model. 

Let us also note that, recently, we have also studied the chaotic dynamics emerging from the mass deformed ABJM model using similar methods and found that the Lyapunov exponents vary either  $\lambda_L \propto (E/N^2)^{1/3}$ or $\lambda_L \propto (E/N^2 - \gamma_N)^{1/3}$, where $\gamma_N(k, \mu)$ is a constant determined in terms of the parameters of this model. Our results, including the critical temperature in regard to the MSS bound are presented in the article \cite{Baskan:2022dys}.

The natural next step appears to be going beyond the classical description and devise means of incorporating the quantum effects given that any direct method of exploration of the full quantum dynamics involving real-time techniques is not presently available. Recently a real-time method\footnote{Most of the earlier investigations, as well as some recent studies \cite{Kawahara:2006hs, Kawahara:2007fn, DelgadilloBlando:2007vx, DelgadilloBlando:2008vi, Asano:2018nol} have been aimed at investigating the phase structure of these models in the Euclidean time formulation using both analytical and Monte-Carlo methods.} involving an approximation using Gaussian states is proposed and thoroughly applied to the BFSS model \cite{Buividovich:2018scl,Buividovich:2017kfk}. This approach incorporates the quantum corrections by considering a larger but a truncated set of observables and the Heisenberg equations of motion are obtained upon using a Gaussian density matrix and extensively employing the Wick's theorem. Results reported in \cite{Buividovich:2018scl} indicate that the Lyapunov exponents vanish at finite temperature implying that the quantum description of the BFSS model within this approximation is in agreement with the MSS inequality. However, it falls short in providing an explicit saturation of the MSS bound by the Largest Lyapunov exponent, in contrast to the result for the SYK model obtained in \cite{Maldacena:2016hyu} and expected from all models with holographic duals according to the MSS conjecture. 

We think that it may be worthwhile to apply this approach to the family of models reported in this talk to test and expand its validity in a broader sense as well as to further explore the chaotic dynamics of the mass deformed matrix gauge theories beyond the classical regime. We hope to report on the new developments along this direction elsewhere. 

\acknowledgments{K.B.,S.K., C.T., acknowledge the support of TUBITAK under the project number 118F100 and the METU research project GAP-105-2018-2809.}

\end{document}